\documentstyle[preprint,version2,aps]{revtex}

\begin{document}
\draft

\begin{title}
\vskip2.5cm
Bosonization approach to \\
the one-dimensional Kondo lattice model
\end{title}
\vskip1cm
\author{Satoshi Fujimoto$^{\rm a}$ and Norio Kawakami$^{\rm b}$}
\begin{instit}
Department of Physics, Faculty of Science$^{\rm a}$, and
Yukawa Institute for Theoretical Physics$^{\rm b}$,\\
Kyoto University, Kyoto 606, Japan
\end{instit}

\begin{abstract}
The one-dimensional Kondo lattice model is investigated by
using bosonization techniques and conformal field theory.
In the half-filled band, the charge and spin gaps open for
the anti-ferromagnetic Kondo coupling.
Away from half-filling, the paramagnetic metallic state
is characterized by the fixed point of Tomonaga-Luttinger
liquid with the {\it large} Fermi surface
in accordance with known results.
It is suggested that as the electron density
approaches half-filling, both of spin and charge susceptibilities
may show a divergence property.
\end{abstract}
\vspace{1 cm}
\narrowtext
\newpage
There is much current interest in the Kondo lattice model (KLM),
which is considered to be a basic model for heavy fermion systems.
The model hamiltonian consists of conduction electrons coupled with
a localized spin array via the Kondo exchange interaction.
The competition between the Kondo effect and the RKKY interaction
results in various phases such as
magnetic phases,  Kondo insulators, etc.
As a first step to understand the KLM, the one-dimensional
(1D) KLM has been studied extensively by
renormalization group methods\cite{pf,car},
exact analytic methods\cite{sig}, numerical
diagonalizations\cite{tsu,ueda}, Monte Carlo
simulations\cite{sca,tro}, and bosonization
methods\cite{strong,tsvelik},
which have clarified basic properties of the model.
In this paper we investigate the low-energy physics of the 1D
KLM by using bosonization and conformal field theory
(CFT) techniques,
and give complementary discussions to the results known so far.
In particular, we point out that the {\it marginally relevant}
spin interaction plays a key role in the model.

We consider the 1D KLM,
\begin{equation}
H=-t\sum_{i,\sigma} c^{\dagger}_{i,\sigma}c_{i+1,\sigma}
+\lambda_K\sum_i \bbox{S}_{c, i}\cdot\bbox{S}_{f, i}
+\lambda_f \sum_i \bbox{S}_{f,i}\cdot\bbox{S}_{f,i+1},
\quad(\lambda_K, \lambda_f>0),
\end{equation}
where the interaction $\lambda_f$ for $f$-spins is introduced,
which makes it easy to treat localized spins by
bosonization methods.
Here $c_{i,\sigma}$ and $c^{\dagger}_{i,\sigma}$,
are the annihilation and
creation operators of conduction electrons, and
$\bbox{S}_c=c^{\dagger}_{i,\alpha}
\bbox{\sigma}_{\alpha,\beta}c_{i,\beta}/2$
with the Pauli matrix $\bbox{\sigma}$.
Similarly, $\bbox{S}_f=f^{\dagger}_{i,\alpha}\bbox{\sigma}_{\alpha,\beta}
f_{i,\beta}/2$ for localized $f$-electrons.
In order to study the low-energy critical
behavior, we use bosonization techniques combined with CFT.
Applying a non-abelian bosonization, we first separate
the charge and spin degrees of freedom of conduction electrons
preserving SU(2) symmetry of the spin sector\cite{wit,aff,aff2}.
This method enables us to observe whether the interaction is
relevant for the gap formation.
In a continuum limit, the electron operators are expressed in
terms of the left-going (L) and right-going (R) operators,
$c_{\sigma}(x)={\it e}^{ik_{\rm F}x}c_{L \sigma}(x)+
{\it e}^{-ik_{\rm F}x}c_{R \sigma}(x).$
Introducing the current operators for the charge
and spin degrees of freedom,
$J_{c,L}=c^{\dagger}_{L\sigma}c_{L\sigma}$,
$\bbox{J}_{c,L}=c^{\dagger}_{L\alpha}\frac{1}{2}
\bbox{\sigma}_{\alpha\beta}
c_{L\beta}$, etc.,
we can represent the hamiltonian by these currents,
\begin{eqnarray}
H&=&H_c+H_s, \nonumber \\
H_c&=&\frac{\pi v}{2} \int{\rm d}x[J_{c,L}(x)J_{c,L}(x)+ J_{c,R}(x) J_{c,R}(x)
 ], \nonumber \\
H_s&=&\frac{2\pi v_c}{3} \int{\rm d}x
[\bbox{J}_{c,L}(x)\cdot\bbox{J}_{c,L}(x)+\bbox{J}_{c,R}(x)
\cdot\bbox{J}_{c,R}(x)]
+\lambda_K\int{\rm d}x\bbox{S}_c(x)\cdot\bbox{S}_f(x) \nonumber \\
&+&\frac{2\pi v_f}{3} \int{\rm d}x
[\bbox{J}_{f,L}(x)\cdot\bbox{J}_{f,L}(x)
+\bbox{J}_{f,R}(x)\cdot\bbox{J}_{f,R}(x)]
, \label{eqn:effh}
\end{eqnarray}
where $v_c$($v$) and $v_f$ are the velocities of the spin
(charge) excitation of conduction electrons and
localized $f$-electrons. $J_{c,L(R)}$ and $\bbox{J}_{c(f),L(R)}$
satisfy U(1) current algebra and level-1 SU(2) current
algebra \cite{wit,aff,aff2}. From non-abelian bosonization formulas,
$c^{\dagger}_{L\alpha}c_{R\beta}
\propto (g_c)_{\alpha\beta}{\it e}^{i\sqrt{2\pi}\phi_c^{(c)}},
$ etc., we can express the spin operator as
\begin{equation}
\bbox{S}_c=\bbox{J}_{c,L}+\bbox{J}_{c,R}
+{\rm const.}({\it e}^{2i k_{\rm F}x}\frac{1}{2}{\rm tr}
(g_c\bbox{\sigma}){\it e}^{i\sqrt{2\pi} \phi_c^{(c)}}+h.c.),
\end{equation}
where $g_c$ is the fundamental representation of SU(2)$\times$SU(2),
and $\phi_c^{(c)}$ is the bosonic phase field related to the charge
degrees of freedom. A similar formula holds for the localized spin
$\bbox{S}_f$ with the charge part $\exp{(i\sqrt{2\pi}\phi_f^{(c)})}$
being replaced by its expectation value.

Let us begin with the half-filled case
in which the number of conduction electrons equals that of
lattice sites. In this case, it has been deduced numerically
that the excitation gaps open both
in the charge and spin sectors characterizing
the Kondo insulator \cite{tsu}. We shall
discuss this analytically below
(see also \cite{tsvelik}). The Kondo interaction
in eq.(\ref{eqn:effh}) is
rewritten in terms of boson fields,
\begin{eqnarray}
H_{int}&=&\lambda_K\int{\rm d}x
(\bbox{J}_{c,L}(x)+\bbox{J}_{c,R}(x))
(\bbox{J}_{f,L}(x)+\bbox{J}_{f,R}(x)) \nonumber \\
&+&\frac{\lambda_K}{4}{\rm const.}
\int{\rm d}x
[{\rm tr}(g_c\bbox{\sigma}){\rm tr}(g_f\bbox{\sigma})
{\it e}^{i\sqrt{2\pi} \phi_c^{(c)}}
+{\rm tr}(g_c\bbox{\sigma}){\rm tr}(g_f^{\dagger}\bbox{\sigma})
{\it e}^{i\sqrt{2\pi} \phi_c^{(c)}}
+h.c.], \label{eqn:int}
\end{eqnarray}
where we have dropped the irrelevant  oscillating terms.
Before switching on the exchange interaction,
the scaling dimensions of the fields
$\exp(i\sqrt{2\pi}\phi_c)$, $g_c$, and $g_f$
are all equal to $1/2$\cite{kz,zam1}.
Therefore the second term of eq.(\ref{eqn:int})
turns out to be a {\it relevant} operator which
has the dimension smaller than $2$.
In order to see in which mode the relevant interaction
opens the mass gap, we rewrite the interaction in  abelian
boson representation. From the formulas,
$J^z_{c,R(L)}\propto\partial_{+(-)}\phi_c,$
$J^{\pm}_{c,R(L)}\propto{\it e}^{\mp i\sqrt{8\pi}\phi_{c,R(L)}},
$ etc., and
\begin{equation}
g_c=\left(
\begin{array}{cc}
\exp (i\sqrt{2\pi}\phi_c)& \exp (i\sqrt{2\pi}\tilde{\phi_c}) \\
\exp (-i\sqrt{2\pi}\tilde{\phi_c})& \exp (-i\sqrt{2\pi}\phi_c)
\end{array}
\right), {\rm etc.},
\end{equation}
where $\phi_c=\phi_{c,L}+\phi_{c,R}$ and
$\tilde{\phi_c}=\phi_{c,L}-\phi_{c,R}$\cite{aff2},
the second term of eq.(\ref{eqn:int}) becomes
$16\lambda_K\cos \sqrt{2\pi}(\tilde{\phi_c}-\tilde{\phi_f})
\cos\sqrt{2\pi}\phi_c^{(c)}$.
Hence, the above relevant interaction may open excitation gaps in
the spin mode
$\psi\equiv(\tilde{\phi_c}-\tilde{\phi_f})/\sqrt{2}$ as
well as in the charge mode $\phi_c^{(c)}$.
After integrating out the massive charge mode, $\phi_c^{(c)}$,
we have the massive $\psi$ mode
and
the massless $\chi\equiv(\tilde \phi_c+ \tilde \phi_f)/\sqrt{2}$ mode
in the spin sector.
This can be easily checked for $v_c=v_f$ where
 the $\phi$ mode and the $\chi$ mode are decoupled.
In the case with $v_c>v_f$, by taking into account that
$v_f$ grows large toward the value $v_f=v_c$ by
the renormalization, one can expect that
the $\chi$ mode is decoupled at the fixed point.
Then, what happens for this massless spin mode $\chi$ if
we take into account  the first term of eq.(\ref{eqn:int})?
Notice  that this interaction is
{\it marginal} with the scaling dimension equal to $2$.
So, it is crucial whether it is
 marginally relevant or irrelevant.
To see this, let us separate the interaction into two parts,
\begin{equation}
H_{int}^m+H_{int}^r=\lambda_K^m(\bbox{J}_{c,L}\cdot\bbox{J}_{f,L}
+\bbox{J}_{c,R}\cdot\bbox{J}_{f,R})
+\lambda_K^r(\bbox{J}_{c,L}\cdot\bbox{J}_{f,R}
+\bbox{J}_{c,R}\cdot\bbox{J}_{f,L}).
\label{eqn:marg}
\end{equation}
The scaling equations for the couplings
$\lambda_K^m$ and $\lambda_K^r$ can be obtained
by expanding the partition function
$Z={\rm tr}{\it e}^{-\beta H}$
in the Kondo coupling and by using the operator product expansion
in SU(2) current algebra\cite{kz,zam1},
\begin{equation}
J^a_{c,L}(z)J^b_{c,L}(z^{'})=
\frac{\varepsilon_{abc}J^c_{c,L}(z^{'})}{2\pi (z-z^{'})}
+\frac{\delta_{ab}}{4\pi^2(z-z^{'})^2}, \hskip 3mm
{\rm etc.}\label{eqn:ope}
\end{equation}
Up to the second order, the scaling equations read
\begin{equation}
\frac{{\rm d}\lambda_K^m}{{\rm d}{\rm ln}L}=0, \qquad
\frac{{\rm d}\lambda_K^r}{{\rm d}{\rm ln}L}
=\frac{(\lambda_K^r)^2}{2\pi(v_c+v_f)}.\label{eqn:scal}
\end{equation}
We can see from (\ref{eqn:scal}) that the
effect of $H^m_{int}$ is just to renormalize velocities, whereas
$H_{int}^r$ is {\it marginally relevant}, causing a spin gap.
An important consequence of the marginal interaction is
that the gap is given by the form of
$\Delta_s\sim\exp (-b/\lambda_K)$ ($b>0$),
which is characterized by a Kosterlitz-Thouless-type transition.
Therefore both of the charge and spin gaps open at half-filling,
describing the fixed point of the  Kondo insulator.
These results agree with those
of the numerical diagonalization
and quantum Monte Carlo studies\cite{tsu,sca}.
In particular, one can see that the spin gap of
$\Delta_s\sim\exp (-b/\lambda_K)$ deduced by numerical
results \cite{tsu} reflect the marginally relevant interaction
in (\ref{eqn:marg})\cite{tsvelik,pf}.

We now turn to the case away from half-filling.
Dropping irrelevant terms including oscillating factors
in eq.(\ref{eqn:int}), we end up with the interaction,
\begin{equation}
H_{int}=\lambda_K \int{\rm d}x
(\bbox{J}_{c,L}(x)+\bbox{J}_{c,R}(x))\cdot\bbox{S_{f}}(x),
\label{eqn:margi}
\end{equation}
which  is still marginally relevant for
the spin sector. In this case, charge excitations become massless,
leading to a metallic state with the renormalized
charge velocity. In order to see properties of the spin part,
we first derive the scaling equation for the Kondo
coupling $\lambda_K$
which is obtained similarly as in the half-filled case,
\begin{equation}
\frac{{\rm d}\lambda_K}{{\rm d}{\rm ln}L}
=\frac{(\lambda_K)^2}{2\pi v_c}.
\label{eqn:scal2}
\end{equation}
We note that Caron and Bourbonnais also derived
the scaling equations (\ref{eqn:scal2})
by using the Kadanoff-Wilson renormalization group method
\cite{car}.
In the infrared limit the Kondo coupling
grows large towards the strong-coupling fixed point.
We recall here that although in
the $\lambda_K \rightarrow \infty$ limit
the ferromagnetic state appears\cite{sig},
a paramagnetic state may be stabilized for
a large parameter space in the metallic phase\cite{tsu2}.
Based on this observation,
we assume the finite value of $\lambda_K^{*}$
for strong-coupling fixed point in the normal metallic phase.
A simple physical picture at this fixed point
is that the marginally relevant interaction
(\ref{eqn:margi}) couples conduction electrons
strongly with $f$-electrons to make singlet clouds,
and remaining unpaired $f$-electrons form a massless spin
mode which carries SU(2) currents.  Strictly speaking,
conduction electrons also contribute to the
massless spin mode because of
the finite value of  $\lambda_K^{*}$ at the fixed point
(see also discussions for the Fermi surface below).
Therefore in the infrared limit, we have
the effective hamiltonian
\begin{equation}
H^{*}=\frac{\pi v^{(c)}}{2}\int{\rm d}x[\{J_{c,L}(x)\}^2+\{J_{c,R}(x)\}^2]
+\frac{2\pi v_f^{'}}{3}
\int{\rm d}x[\bbox{J}_{f,L}^{'}(x)\cdot\bbox{J}_{f,L}^{'}(x)
+\bbox{J}_{f,R}^{'}(x)\cdot\bbox{J}_{f,R}^{'}(x)], \label{eqn:fixh}
\end{equation}
where $v^{(c)}$ and $v_f^{'}$ are the renormalized velocities
of the charge and spin excitations. Note
that the new SU(2) spin current
$\bbox{J}_{f,L(R)}^{'}$ include both of
conduction electrons and $f$-electrons.
The above hamiltonian consists of the massless charge mode
(holon) described by U(1) gaussian CFT and the massless spin mode
(spinon) described by level-1 SU(2) CFT\cite{zam1}.
Hence the fixed point in the metallic phase
belongs to the universality class of
the Tomonaga-Luttinger (TL) liquid
as for the Hubbard model\cite{ueda,hal,os,fk,ky,sh}.
 Since this result is consistent with
the conclusion deduced by Ueda et al. for $\lambda_f=0$
\cite{ueda}, we believe that the introduction of
the interaction $\lambda_f$ in eq.(1) may not
change the essential physics of the Kondo lattice model
even in the case away from half-filling.

Now we wish to ask  how is the volume of the Fermi surface?
In ref. \cite{ueda}, it has been claimed to be {\it large}.
In the bosonization language, the Fermi surface may be
described as follows.
Since there is no relevant interaction in the charge sector,
the left-going and right-going currents
are decoupled and backward scatterings become irrelevant
in the infrared limit.
Thus the effect of the Kondo interaction on the charge sector
at the strong coupling fixed point is just to give rise the $\pi/2$
phase shift\cite{phase}.
The total phase shift of conduction electrons due to
$N_f$ localized spins is $\pi N_f/2$, which
changes the "pseudo-Fermi surface" of holons from
$2k_{\rm F}$ to $2k_{\rm F}+\pi$\cite{chfs}. This argument is not
sufficient  to deduce the real Fermi surface.
As for the spin sector, the marginally relevant
interaction (\ref{eqn:margi}) again plays a crucial role
to hybridize two kinds of spinons consisting of conduction electrons
(pseudo-Fermi surface $k_{\rm F}$) and
$f$-electrons ($\pi/2$), making the new "pseudo-Fermi surface" of
massless spinons as $k_{\rm F} + \pi/2$.
Note that these pseudo-Fermi surfaces of
holons and spinons are consistent with the formation of
the charge and spin gaps at half-filling $k_{\rm F}=\pi/2$.
Combining these pseudo-Fermi
surfaces, we can say that not only conduction
electrons but also localized
electrons contribute to the Fermi surface,
forming the {\it large Fermi surface}, $k_{\rm F}+\pi/2$,
as we shall see more directly in the arguments for the single-particle
Green function mentioned below.
The result is in accordance with the conclusion of
Ueda et al.\cite{ueda}.

Let us now briefly discuss the critical exponents of
correlation functions
in the normal metallic phase. According to the above discussions,
we have two kinds of primary fields
$\Phi^h_{\Delta^{\pm}_h}(x)$ (holon) and
$\Phi^s_{\Delta^{\pm}_s}(x)$ (spinon)
with  conformal  dimensions \cite{fk,ky},
\begin{eqnarray}
\Delta_h^{\pm}&=&\biggl(\sqrt{K_{\rho}}\biggl(D_c+\frac{D_s}{2}\biggr)
\pm \frac{I_c}{4\sqrt{K_{\rho}}}\biggr)^2  \nonumber \\
\Delta_s^{\pm}&=&\frac{1}{4}\biggl(I_s-\frac{I_c}{2}\pm D_s \biggr)^2,
\end{eqnarray}
which satisfy the requirements of U(1) and SU(2)
current-algebra symmetries respectively.
Here $I_c$ ($I_s$) and $D_c$ ($D_s$) are quantum numbers for
the charge (spin) degrees of freedom, which
obey the selection rules of
Fermi statistics, $D_c=(I_c+I_s)/2$ and $D_s=I_c/2$
mod 1 \cite{fk,ky}.
$K_{\rho}$ is a parameter which features U(1) critical line,
 depending on the bare parameters
$\lambda_K/t$ and $\lambda_f/t$.
The operators of conduction- and $f$-electrons can be
expanded in terms of these primary fields as
$c_{\sigma}(x)=\sum_{\Delta_h^{\pm}\Delta_s^{\pm}}
a_{\Delta_h^{\pm} \Delta_s^{\pm}}
\Phi^h_{\Delta^{\pm}_h}(x)\Phi^s_{\Delta^{\pm}_s}(x)$, etc,
and hence various  correlation functions in the asymptotic region
are determined by those for the primary fields\cite{fk,ky}.
For example,
 the single particle Green function
for the conduction electrons,
$\langle c^{\dagger}_{\sigma}(x)c_{\sigma}(0)\rangle$,
is obtained by taking the quantum numbers
$(I_c,I_s,D_c,D_s)=(1,1,0,1/2)$.
The resulting phase factor $\exp(2iD_c k_h x)\exp(2iD_s k_s x)$,
where $k_h$ and $k_s$ are
the ``pseudo-Fermi surfaces'' for holon and spinon
given above, determines the position of the
singularity in the Green function, i.e., the Fermi surface.
Thus $\langle c^{\dagger}_{\sigma}(x)c_{\sigma}(0)\rangle$ shows
the singularity at $k_{\rm F}+\pi/2$ (large Fermi surface)
with the exponent $\eta =(K_{\rho}+1)^2/4K_{\rho}$.
Next we consider spin-spin correlation functions,
$\langle S_c^{+}(x,t)S_c^{-}(0,0)\rangle$,
$\langle S_f^{+}(x,t)S_f^{-}(0,0)\rangle$, and
$\langle S_c^{+}(x,t)S_f^{-}(0,0)\rangle$,
which
show power-law decay with the same exponents determined by the primary
field with $(I_c, I_s, D_c, D_s)=(0,0,1,-1/2)$
such that
\begin{equation}
\langle S_f^{+}(x,0)S_f^{-}(0,0)\rangle
\sim \frac{A_0}{x^2}+\frac{A_1}{x^{\alpha}}(-1)^x \cos 2k_{\rm F}x,
\hskip 3mm {\rm etc.},
\label{eqn:ff}
\end{equation}
where $\alpha=1+K_{\rho}$. We can deduce the
correlation exponents in some limiting cases.
For the weak coupling limit ($\lambda_K \rightarrow 0$)
the critical exponent may be characterized by
that of free electrons with $K_{\rho} \simeq 1$.
Near half-filling, the charge sector
scales to the strong correlation limit
characterized by the spinless-fermion exponent $K_{\rho}=1/2$.
It is hence predicted that the spin exponent
near half-filling takes $\alpha=3/2$
and that for the momentum distribution is $\theta=1/8$,
for any value of the Kondo coupling $\lambda_K$.
In general the correlation exponents may
range $1/2\leq K_{\rho}\leq 1$, similarly to  the case of the
Hubbard model\cite{os,fk,ky,sh}.
It is instructive to note that the $f$-spin correlation
(\ref{eqn:ff}) shows a power-law
with $2k_{\rm F}+\pi$ oscillation (neither $2k_{\rm F}$
nor $\pi$).
This long-distance behavior may be
observed in the energy scale smaller than
$\sim \exp(-b/\lambda_K)$,
although the RKKY interaction
may control the short-distance behavior,
making characteristic structures around  $q=2k_{\rm F}$
in correlation functions \cite{shiba}.
In particular, for small $\lambda_K$
the $q=2k_{\rm F}$ structure may become more prominent
\cite{tsu,yu}. Even in this case, however,
the low-energy critical behavior may be determined by
the $2k_{\rm F}+\pi$ singularity so long
as $\lambda_K$ is finite.

Finally some comments are in order for bulk quantities.
For example, one can expect the spin susceptibility to
show a logarithmic low-temperature dependence in the
metallic phase due to the leading irrelevant (marginal)
operator such as $\bbox{J}_{f,L}^{'}\cdot\bbox{J}_{f,R}^{'}$,
as observed in $s=1/2$ Heisenberg chain\cite{aff3},
while it may exhibit exponential dependence at half-filling.
Another remarkable feature is that
as the electron density approaches half-filling
at zero temperature, the
charge susceptibility as well as the spin susceptibility may exhibit
a divergence property reflecting the formation
of the gaps, as observed for
the charge susceptibility of the
1D Hubbard model\cite{kawa}.  Particularly,
the spin susceptibility is predicted to behave like
$\chi_s \simeq \exp (-b/\lambda_K)(1-n)^{-1}$ for small $\lambda_K$
reflecting the marginally relevant interaction
($n$: density of conduction electrons).

In this paper, we have not discussed
the effects of $\lambda_f$ in detail, which
may appear through the renormalization of the velocity $v_f$.
The scaling equation for  $\lambda_f$
has a similar form to eq.(\ref{eqn:scal2}).
So, there is a possibility that a massless spin phase
may be stabilized at half-filling  due to the large
renormalization effect of $v_f$,
if $\lambda_f$  exceeds
a certain critical value $\lambda_c$ initially. We think that
the results obtained here, such as the large Fermi surface,
are valid in the region for small $\lambda_f$ ($<\lambda_c$).

We have mainly discussed the case with the
antiferromagnetic Kondo coupling.
In the ferromagnetic Kondo case, it is
known that a gap (\'a la Haldane gap) opens [4].
Unfortunately our formalism
cannot describe the gap formation for this case.
We think that another relevant mechanism which may be
dropped in our formalism
should be taken into account for generating the gap for the
ferromagnetic case. This point should
be clarified in the future study.

This work was partly supported by a Grant-in-Aid from the Ministry
of Education, Science and Culture.

\newpage


\begin{references}
\bibitem{chfs}
In the reduced Brillouin zone, the fermi surface for holons is
$\pi-2k_{\rm F}$.
\end{references}
\end{document}